\documentstyle[11pt,AATS,epsf]{article}	
\begin{document}	

\title{High Redshift Quasars and Star Formation History}

\author{M. Dietrich and F. Hamann}
\affil{University of Florida, Department of Astronomy, 211 Bryant Space 
       Science Center, Gainesville, FL 32611-2055, USA}

\begin{abstract}
Quasars are among the most luminous objects in the universe, and they can 
be studied in detail up to the highest known redshift.
Assuming that the gas associated with quasars is closely related to the 
interstellar medium of the host galaxy, quasars can be used as tracer of 
the star formation history in the early universe.
We have observed a small sample of quasars at redshifts $3\la z \la 5$ and
present results using NV/CIV and NV/HeII as well as MgII/FeII to estimate 
the date of the first major star formation epoch. 
These line ratios indicate solar and supersolar metallicities of the gas 
close to the quasars. Assuming times of $\tau _{evol} \simeq 1$\,Gyr
the first star formation epoch can be dated to $z_f \simeq 10$, 
corresponding to an age of the universe of less than $5\cdot 10^8$\,yrs 
(H$_{\rm o} = 65$\,km\,s$^{-1}$\,Mpc$^{-1}$, $\Omega_M$=0.3, 
 $\Omega _\Lambda = 0.7$).
\end{abstract}
\section{Introduction}
In the context of cosmic evolution, the epoch of first star formation in the 
early universe is of fundamental importance.
During the last few years, several galaxies 
(cf., Dey et al.\,1998; Weymann et al.\,1998; Spinrad et al.\,1998;
Chen et al.\,1999; van Breugel et al.\,1999; Hu et al.\,1999)
and quasars 
(Fan et al.\,1999, 2000a, 2000b; Zheng et al.\,2000; Stern et al.\,2000)
at redshifts of $z\geq 5$ have been detected.
Because quasars are among the most luminous objects in the universe, they 
are valuable probes of conditions at early cosmic times. 
One particularly important diagnostic is their gas metallicity. 
If the gas near high redshift quasars is related to the interstellar matter 
of the young host galaxies, quasars can be used to probe the star 
formation and chemical enrichment history of those galactic environments.
Recent studies of quasars at moderately high redshifts ($z\ga3 $) show solar 
and enhanced metallicities in the line emitting gas 
(cf., Hamann \& Ferland 1993; Osmer et al.\,1994; Ferland et al.\,1996; 
Hamann \& Ferland 1999; Dietrich \& Wilhelm-Erkens 2000).
These results require a rapid and efficient phase of star formation in the 
early universe, e.g.\, in the dense galactic or proto-galactic nuclei where 
quasars reside.

In the following, we present results of an ongoing study of quasars at 
redshifts $3 \la z \la 5$. The emission line ratios of NV1240 to CIV1549 and 
HeII1640 are used as well as MgII2798 vs.\,FeII\,UV. The relative strength of 
these ratios indicates that the first epoch of star formation started at 
redshifts $z\geq 10$.
In current cosmological models, the age of the universe at those redshifts 
is less than $5\cdot 10^8$\,yrs 
(H$_{\rm o}$=65\,km\,s$^{-1}$ Mpc$^{-1}$, $\Omega _M $=0.3, 
 $\Omega _\Lambda$=0.7; cf., Carroll et al.\,1992).
\section{Observations}
The observations of high redshift quasars were carried out at several 
observatories during 1993 and 2000. We used telescopes at Calar Alto
Observatory/Spain, McDonald Observatory/Texas, USA, 
La Silla Observatory/ESO,Chile, Paranal Observatory/ESO,Chile, Keck/Hawaii,USA,
and CTIO/Chile (Tab.\,1).
\begin{table}		
\begin{center}
\caption{Observing log of the studied high redshift quasars}
\vspace*{0.5cm}		
\begin{tabular}{lclcl}	
\tableline
quasar  &$z$&observatory&$\lambda \lambda $-range [\AA ]&date\\
\hline	
\hline	
UM\,196     &2.81&Calar Alto, 3.5m&3800-8200  &Aug.\,1993\\
BRI\,0019-1522&4.52&CTIO, 4m      &11500-23600&Sept.\,2000\\
Q\,0044-273 &3.16&Paranal, 8.2m   &3800-9400  &Jul.\,1999\\
UM\,667     &3.13&Calar Alto, 3.5m&3800-8200  &Aug.\,1993\\
Q\,0046-282 &3.83&Paranal, 8.2m   &3800-9400  &Jul.\,1999\\
Q\,0103+0032&4.44&CTIO, 4m      &11500-23600  &Sept.\,2000\\
Q\,0103-294 &3.12&Paranal, 8.2m   &3800-9400  &Jul.\,1999\\
Q\,0103-260 &3.36&Paranal, 8.2m   &3800-9400  &Jul.\,1999\\
            &    &La Silla, 3.5m  &9500-24800 &Oct.\,1999\\
Q\,0105-2634&3.48&La Silla, 3.5m  &9500-24800 &Oct.\,1999\\
4C\,29.05   &2.36&Calar Alto, 3.5m&3800-8200  &Aug.\,1993\\
Q\,0216+0803&2.99&McDonald, 2.7m  &3800-7800  &Jul.\,1995\\
PSS\,J0248+1802&4.44&CTIO, 4m     &11500-23600&Sept.\,2000\\
Q\,0256-0000&3.37&La Silla, 3.5m  &9500-24800 &Oct.\,1999\\
Q\,0302-0019&3.29&La Silla, 3.5m  &9500-24800 &Oct.\,1999\\
PC\,1158+4635&4.73&Keck, 10m      &12700-24700&May\,2000\\
HS\,1425+60 &3.19&Calar Alto, 3.5m&3800-8200  &Aug.\,1993\\
Q\,1548+0917&2.75&McDonald, 2.7m  &3800-7800  &Jul.\,1995\\
PC\,1640+4711&2.77&McDonald, 2.7m  &3800-7800 &Jul.\,1995\\
HS1700+64   &2.74&Calar Alto, 3.5m&3800-8200  &Aug.\,1993\\
PKS\,2126-15&3.28&Calar Alto, 3.5m&3800-8200  &Aug.\,1993\\
PC\,2132+0216&3.19&Calar Alto, 3.5m&3800-8200  &Aug.\,1993\\
Q\,2227-3928&3.44&La Silla, 3.5m  &9500-24800 &Oct.\,1999\\
Q\,2231-0015&3.02&Calar Alto, 3.5m&3800-8200  &Aug.\,1993\\
            &    &McDonald, 2.7m  &3800-7800  &Jul.\,1995\\
BRI\,2237-0607&4.57&CTIO, 4m      &11500-23600&Sept.\,2000\\
UM\,659     &3.04&Calar Alto, 3.5m&3800-8200  &Aug.\,1993\\
Q\,2348-4025&3.31&La Silla, 3.5m  &9500-24800 &Oct.\,1999\\
\tableline \tableline
\end{tabular}
\end{center}
\end{table}

The redshift range of $2.7\la z \la 3.3$ was chosen to assure that most of the
diagnostic ultraviolet lines are shifted into the optical regime, in particular
the NV1240, CIV1549, and HeII1640 emission lines.
The quasars which we observed in the near infrared domain 
($\sim 1 - 2.5 \mu$m) were selected for their brightness and for a suitable
redshift ($3.3\la z \la 4.7$) that the MgII2798 and the broad FeII emission 
features in the ultraviolet were shifted to the J- or H-band, respectively.

\subsection{The Method}
Quasars show a prominent emission line spectrum which provides information 
on the physical conditions of the gas i.e.\,temperature, density, ionization 
state, and the chemical composition.
Although the ratios of strong emission lines like Ly$\alpha$1215 to CIV1549 
are quite insensitive to the metallicity, other ratios can provide indirect 
constraints.

The key to using emission line ratios to estimate the metallicity is the
different production rates of primary elements like carbon and secondary 
elements, like nitrogen. 
N is selectively enhanced by secondary processing at moderate to high 
metallicities, leading to N increasing as roughly Z$^2$
(cf., Hamann \& Ferland 1993; Vila-Costas \& Edmunds 1993).
Recent model calculations provide evidence for a strong metallicity 
dependence of emission line ratios involving such elements. Hence,
NV1240 vs.\,CIV1549 and NV1240 vs.\,HeII1640 are of particular interest 
for determining the chemical composition of the gas 
(cf., Hamann \& Ferland 1999 for a review).

The different time scales of the enrichment of gas with ``$\alpha$-elements''
(e.g., O and Mg) and iron are another important aspect using emission line 
ratios to probe the star formation history.
$\alpha $-elements are produced predominantly in massive stars on short 
time scales. These elements are released from massive-star supernovae 
(Types II,Ib,Ic).
The dominant source of iron is ascribed to intermediate mass stars in 
binary systems ending in supernova type Ia explosions 
(cf., Wheeler et al.\,1989).
The amount of iron returned to the interstellar medium in SN\,II ejecta is 
rather low (e.g., Yoshii et al.\,1998).
The significantly different time scales of the release of $\alpha$-elements
and iron to the interstellar medium results in a time delay of the order
of $\sim 1$\,Gyr.
Detecting strong FeII emission at high redshift can be taken as an 
indication that the star formation of the stars which had released the iron 
had occurred $\sim 1$\,Gyr earlier.
The viability of the FeII/MgII emission line ratio as an abundance indicator 
was discussed by Hamann \& Ferland (1999).

\section{Results}
\subsection{NV1240 vs. CIV1549 and HeII1640 Line Ratios}
The quasars observed in the optical wavelength range were used 
to determine the NV1240/CIV1549 and NV1240/HeII1640 emission line ratios.
To measure the NV1240 line strength we had to deblend the Ly$\alpha$1215, 
NV1240 emission line complex.  
We also deblended the CIV1549, HeII1640, OIII]1663 emission line complex 
to measure HeII1640 (cf., Dietrich \& Wilhelm-Erkens 2000 for more details 
of the deblending).
The measured line ratios of NV1240/CIV1549 and NV1240/HeII1640 are compared
to theoretical predictions (Fig.\,1). 
Both line ratios are in good agreement with results obtained by Hamann \& 
Ferland (1992, 1993) for quasars at similar redshift.
The measured line ratios were used to calculate an average line ratio 
yielding NV1240/CIV1549\,=\,0.7$\pm$0.3 and 
NV1240 /HeII1640\,=\,5.9$\pm$3.6.
The dotted lines in Fig.\,1 indicate the line ratios expected for {\it typical}
conditions of the broad emission line region (BELR) assuming solar 
metallicities (Hamann \& Ferland 1999). The observed line ratios are obviously
larger than those for solar metallicities indicating super-solar abundances.

\begin{figure}[]	
\plotfiddle{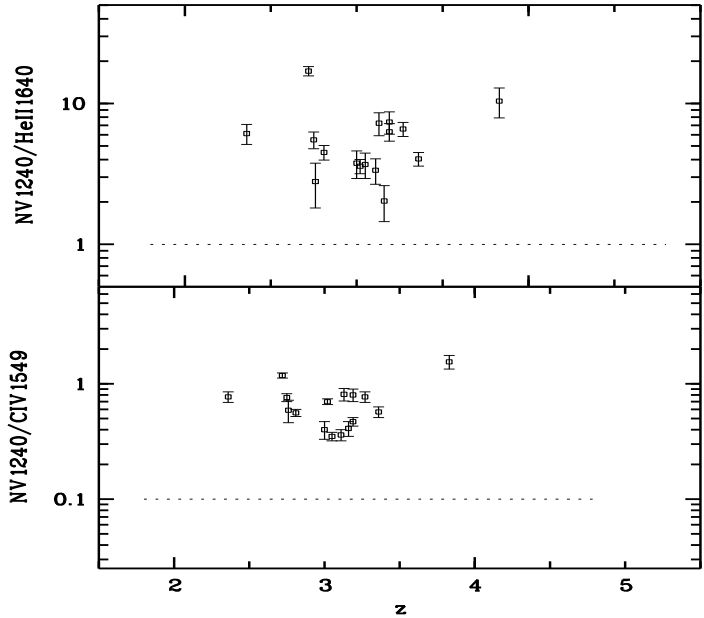}{7.7cm}{0}{100}{75}{-190}{-175}
\caption{NV1240/CIV1549 and NV1240/HeII1640 as a function of redshift.
         The dotted lines indicate the line ratios which are expected 
         for typical conditions of the BELR gas assuming solar metallicity.} 
\end{figure}

\noindent
The conversion of observed emission line ratios to relative abundances is 
affected by several uncertainties. One has to consider only lines, such as
NV1240, CIV1549, and HeII1240, that originate in the same region of the 
BELR under comparable conditions of the gas. 
A detailed discussion of the current limitations of the method can be found in 
Baldwin et al.\,(1996), Ferland et al.\,(1996), or Hamann \& Ferland (1999).
Our abundance estimates are based on the model calculations presented  by 
Hamann \& Ferland (1992, 1993). 
They computed abundances for a large range of evolutionary scenarios and input
the results into numerical models of the BELR.
They varied the slope of the IMF, the evolutionary time scale for the star 
formation, as well as the low mass cutoff of the IMF.
They concluded that the high metallicities observed in high redshift quasars 
can be achieved only in models with rapid star formation (RSF) and a shallow
IMF (slightly favoring massive stars compared to the solar neighborhood), 
comparable to models of giant elliptical galaxies.
It is reassuring that the rapid star formation scenario indicates the same 
range of metallicities based on both the NV1240/CIV1549 and NV1240/HeII1640
line ratios. 
We estimated an abundance of Z\,$\simeq$\,8$\pm$4\,Z$_\odot$ given by our
observed NV/CIV and NV/HeII within the framework of the RSF model
(cf., Dietrich \& Wilhelm-Erkens 2000).

\subsection{MgII2798 vs. FeII\,UV Line Ratio}
The line ratio of $\alpha $-element vs.\,iron emission can be used as a 
cosmological clock because the time scales for the release of $\alpha$-elements
and iron to the interstellar medium are significantly different. The
enrichment delay is of the order of $\sim 1$\,Gyr 
(cf., Wheeler et al.\,1989; Yoshii et al.\,1998).
The best indicator of $\alpha$-

\begin{figure}[]	
\plotfiddle{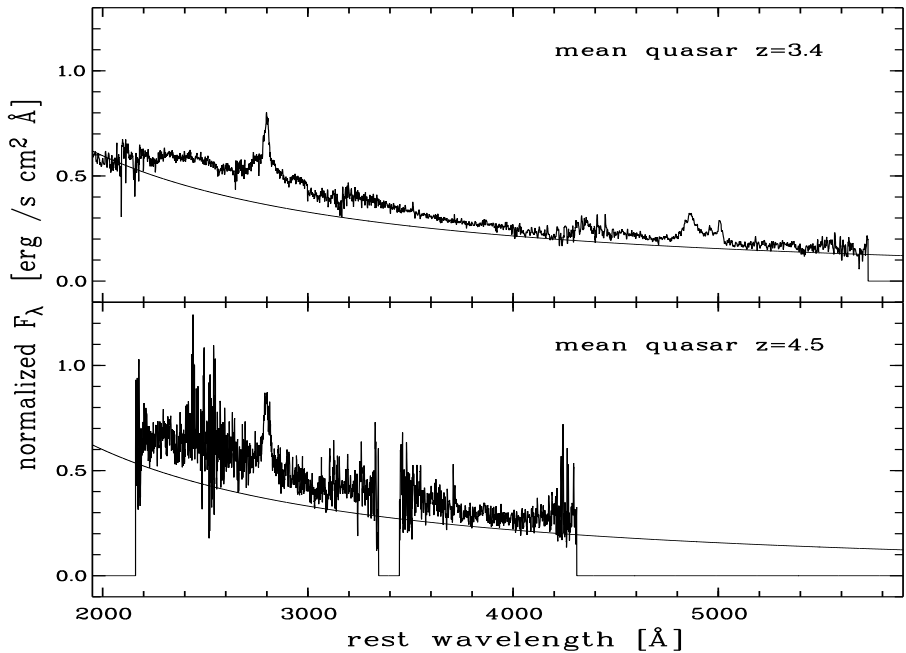}{8.2cm}{0}{70}{49}{-210}{-140}
\caption{The mean quasar spectra at $\overline{z}=3.4$ and
         $\overline{z}=4.5$ together with a powerlaw continuum fit
         with $\alpha = -0.5$ (F$_\nu \sim \nu ^{\alpha}$).}
\end{figure}

\noindent
elements vs.\,iron in quasars is the strength 
of MgII2798 emission compared to broad blends of FeII multiplets spanning 
several hundred \AA ngstroem (rest-frame) on either side of the MgII line 
(cf., Wills et al.\,1980,1985; Zheng \& O'Brien 1990; Boroson \& Green 1992;
Laor 1995; Vestergaard \& Wilkes 2001).

Very few quasars at redshifts larger than $z = 3$ were observed for 
the wavelength region covering MgII2798 to H$\beta$,[OIII]4959,5007
(Hill et al.\,1993; Elston et al.\,1994; Kawara et al.\,1996; 
Taniguchi et al.\,1997; Yoshii et al.\,1998; Murayama et al.\,1998).
Recently, Thompson et al.\,(1999) studied a few quasars at average redshifts
of $\overline{z} = 3.4$ and $\overline{z} = 4.5$, respectively. They 
found no significant difference in the strength of the ultraviolet FeII 
emission relative to MgII2798, which suggests an age of the universe of 
more than 1 Gyr at $z\simeq 4.5$.

In contrast to earlier studies, our data cover a much wider range of
rest frame wavelengths, 
$\lambda \lambda  2100 - 5600$\AA\ and $\lambda \lambda  2100 - 4300$\AA ,
(Fig.\,2). 
These wide and continuous wavelength range enabled us to investigate the 
strong ultraviolet FeII emission based on a 
reliable continuum fit which was hard to achieve in earlier studies with 
smaller and non-continuous wavelength coverage.

Due to the huge number of individual FeII emission lines ($\sim 10^5$) 
it is not practical to treat them individually. 
As suggested and demonstrated by Wills, Netzer, \& Wills (1985), the 
reconstruction of a quasar spectrum by several well defined components, i.e.
    (i)   a power law continuum,
    (ii)  a Balmer continuum emission spectrum,
    (iii) a template for the FeII emission,
and (iv)  a template spectrum for the broad emission lines,
is the best approach to measuring the strength of the FeII emission. 
We are presently involved in a collaboration (cf., Verner et al.\,1999) to 
use state-of-the-art computer models as well as empirical

\begin{figure}[]	
\plotfiddle{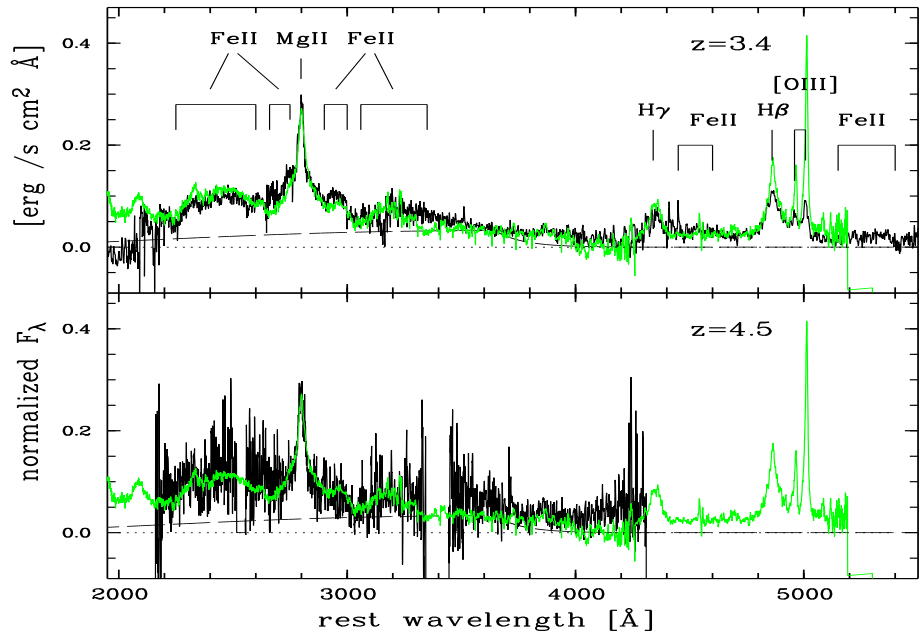}{7.95cm}{0}{70}{47.5}{-210}{-130}
\caption{Comparison of the continuum subtracted mean quasar spectra at 
         $\overline{z}=3.4$ (top) and $\overline{z}=4.5$ (bottom).
         The lightgrey curve shows the local mean quasar spectrum.
         The long dashed line indicates the strength of the Balmer 
         continuum emission.}
\end{figure}

\noindent
FeII,FeIII emission 
templates (Vestergaard \& Wilkes 2001) to quantify the abundance sensitivities
of the FeII line emission.

To obtain a first estimate of the iron emission strength in comparison to
quasars in the local universe, we compared the restframe quasar spectra in our 
samples ($\overline{z}=3.4$ and $\overline{z}=4.5$) to a mean quasar spectrum.
The mean quasar spectrum was calculated from a subset of a large quasar
sample ($>$700 quasars) which we compiled from ground-based 
observations and from archive spectra measured with IUE and HST
(Dietrich \& Hamann 2001).
The mean spectrum used for this comparison is based on 101 quasars with
i) redshift $z\leq 2$, and ii) luminosities in the same range as our
$\overline{z}=3.4$ and $\overline{z}=4.5$ quasars.

The mean quasar spectra of our samples at $\overline{z}\simeq 3.4$ and 
$\overline{z}\simeq 4.5$ are shown in Fig.\,2 together with power law 
continuum fits. 
The continuum fits were subtracted and the pure emission line flux was 
compared. For wavelengths $\lambda \geq 3200$\AA\ much of the emission can 
be attributed to Balmer continuum emission, but for $\lambda \leq 3200$\AA\ 
most of the emission is due to broad FeII emission features (Fig.\,3).
The relative emission strength of the FeII emission of the mean high redshift 
quasars are nearly identical ($\la 15\%$) compared to the mean $z\leq 2$ 
quasar (Fig.\,3).
The mean quasar spectra at $\overline{z}\simeq 3.4$ and 
$\overline{z}\simeq 4.5$ themselves differ by less than $\sim 20$\% . 
This can be taken as an indication for no significant evolution in 
$\alpha $-element vs.\,iron in quasars from the local universe 
to $z\sim 4.5$.

\begin{figure}[]	
\plotfiddle{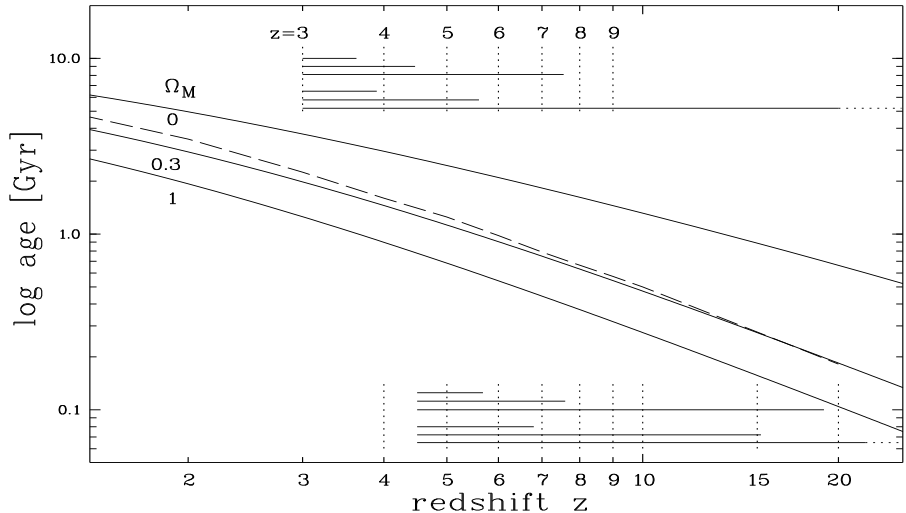}{6.19cm}{-90}{47}{38}{-190}{206}
\caption{Estimate of $z_f$ for several combinations of $\Omega _M$ and 
         $\tau _{evol}$ (H$_o$\,=\,65 km\,s$^{-1}$\,Mpc$^{-1}$).
         The three solid lines show the age of the universe as a function of 
         redshift ($\Omega _M$=0,0.3,1). 
         The long dashed line shows the effect of $\Omega _\Lambda = 0.7$ 
         on the age ($\Omega _M$=0.3).
         The length of the horizontal lines in the upper part of the figure 
         marks $z_f$ for $\tau _{evol}$\,=\,0.5,1,2 Gyrs ($\Omega _M$=0 and
         0.3, respectively).
         The horizontal lines in the lower part of the figure show the same
         but starting now at an average redshift of $z=4.5$.}
\end{figure}

\section{Summary and Discussion}
The presented study of quasars at redshifts $z\simeq 3$ provides evidence for 
higher solar abundances of the line emitting gas. This result is based on 
the emission line ratios of NV1240 vs.\,CIV1549 and HeI1640.
Using assumptions on stellar evolution time scales which are necessary to 
produce solar or higher metallicities, the beginning of the first star 
formation epoch can be estimated.
In Fig.\,4 the age of the Universe is displayed as a function of redshift 
$z$ for several settings of H$_{\rm o}$ and $\Omega _M$.
With an evolutinary time scale of 
$\tau _{evol} \simeq 1$ Gyr ($\Omega _M = 0.3$) or
$\tau _{evol} \simeq 2$ Gyrs ($\Omega _M = 0$) based on normal chemical 
evolution models, the beginning of the first violent star formation episode 
can be dated to a redshift of $z_f \simeq 6-8$ based on 
NV1240/CIV1549 and NV1240/HeII1640 (Fig.\,4).

Assuming an evolutionary time scale of $\tau _{evol} \simeq 1$\,Gyr for the
progenitor stars of type SN\,Ia, we used the MgII2798/FeII UV emission ratio 
as a tracer of star formation history.
The similar MgII/FeII\,UV emission ratios in our high
redshift quasars compared to local quasars suggests an age of the universe of 
$\sim 1$\,Gyr at $z\simeq 4.5$, implying a redshift of $z_f \simeq 8 - 15$ 
($\Omega _M = 0 - 0.3$) for the epoch of the first substantial star formation.
The measured MgII/FeII UV emission ratio probably also suggests at least solar 
abundances.

We concluded, therefore, that high redshift quasars indicate a redshift 
of $z_f \simeq 10$ for the first major star formation epoch, corresponding 
to an age of the universe of $\la 5\cdot 10^8$ yrs 
(H$_o = 65$\,km\,s$^{-1}$\,
Mpc$^{-1}$, $\Omega _M$=0.3, $\Omega _\Lambda = 0.7$).


\acknowledgments
This work was supported by NASA grant NAG\,5-3234 and by the 
Deutsche Forschungsgemeinschaft, project SFB328 and SFB439.
\end{document}